\documentstyle[12pt]{article}
\newcommand{\be}{\begin{equation}}
\newcommand{\ee}{\end{equation}}
\newcommand{\bea}{\begin{eqnarray}}
\newcommand{\eea}{\end{eqnarray}}
\newcommand{\al}{\alpha}

\newcommand{\dl}{\delta}

\newcommand{\zt}{\zeta}
\newcommand{\et}{\eta}
\newcommand{\th}{\theta}
\newcommand{\Th}{\Theta}

\newcommand{\kp}{\kappa}
\newcommand{\lm}{\lambda}
\newcommand{\ks}{\xi}

\newcommand{\ups}{\upsilon}
\newcommand{\ph}{\phi}
\newcommand{\Ph}{\Phi}

\newcommand{\ch}{\chi}
\newcommand{\ps}{\psi}
\newcommand{\Ps}{\Psi}
\newcommand{\om}{\omega}
\newcommand{\Om}{\Omega}

\newcommand{\nn}{\nonumber}
\newcommand{\varep}{\varepsilon}

\begin{document}

\title{{\bf Interactive Quadratic Gravity}}

\author{K Kleidis, A Kuiroukidis and D B Papadopoulos\\
{\small Department of Physics}\\
{\small Section of Astrophysics, Astronomy and Mechanics}\\
{\small Aristotle University of Thessaloniki}\\
{\small 54006 Thessaloniki, Greece} }

\date{}

\maketitle

\begin{abstract}

A quadratic semiclassical theory, regarding the interaction of gravity with a 
massive scalar quantum field, is considered in view of the renormalizable 
energy-momentum tensor in a multi-dimensional curved spacetime. According to 
it, a self-consistent coupling between the square curvature term ${\cal R}^2$ 
and the quantum field $\Ph (t, \vec{x})$ should be introduced in order to 
yield the "correct" renormali-zable energy-momentum tensor in quadratic gravity 
theories. The subsequent interaction discards any higher-order derivative 
terms from the gravitational field equations, but, in the expence, it 
introduces a {\em geometric source} term in the wave equation for the quantum 
field. Unlike the conformal coupling case $(\sim {\cal R} \Ph^2)$, this term 
does not represent an additional "mass" and, therefore, the quantum field 
interacts with gravity not only through its mass (or energy) content $(\sim 
\Ph^2)$, but, also, in a more generic way $({\cal R}^2 \Ph)$. Within this 
context, we propose a general method to obtain mode-solutions for the quantum 
field, by means of the associated Green's function in an anisotropic 
six-dimensional background.

\end{abstract}

\section*{1. Introduction}

In the last few decades there has been a remarkable progress in 
understanding the quantum structure of the non-gravitational fundamental 
interactions (Nanopoulos 1997). On the other hand, so far, there is no quantum 
framework consistent enough to describe gravity itself (Padmanabhan 1989), 
leaving string theory as the most successful attempt towards this direction 
(Green et al 1987, Polchinsky 1998, Schwarz 1999). Within the context of 
General Relativity (GR), one usually resorts to the perturbative approach, 
where string theory predicts corrections to the Einstein equations. Those 
corrections originate from higher-order curvature terms arising in the 
string action, but their exact form is not yet being fully explored 
(Polchinsky 1998).

A self-consistent mathematical background for {\em higher-order gravity 
theories} was formulated by Lovelock (Lovelock 1971). According to it, 
the most general gra-vitational Lagrangian reads
\be
{\cal L} = \sqrt {- g} \sum_{m=0}^{n/2} \lm_m {\cal L}^{(m)}
\ee
where $\lm_m$ are constant coefficients, $n$ denotes the spacetime 
dimensions, $g$ is the determinant of the metric tensor and 
${\cal L}^{(m)}$ are functions of the Riemann curvature tensor 
${\cal R}_{ijkl}$ and its contractions ${\cal R}_{ij}$ and ${\cal R}$, 
of the form 
\be
{\cal L}^{(m)} = {1 \over 2^m} \dl_{i_1...i_{2m}}^{j_1...j_{2m}} 
{\cal R}_{j_1 j_2}^{i_1 i_2} ... {\cal R}_{j_{2m-1} j_{2m}}^{i_{2m-1} i_{2m}}
\ee
where Latin indices refer to the n-dimensional spacetime and 
$\dl_{i_1...i_{2m}}^{j_1...j_{2m}}$ is the generalized Kronecker symbol. 
In Eq. (2), ${\cal L}^{(1)} = {1 \over 2} {\cal R}$ is the Einstein-Hilbert 
(EH) Lagrangian, while ${\cal L}^{(2)}$ is a particular combination of 
quadratic terms, known as the Gauss-Bonnett (GB) combination, since 
in four dimensions it satisfies the functional relation 
\be
{\dl \over \dl g^{\mu \nu}} \int \sqrt {-g} \left ( {\cal R}^2 - 4 
{\cal R}_{\mu \nu}{\cal R}^{\mu \nu} + {\cal R}_{\mu \nu \kp \lm}
{\cal R}^{\mu \nu \kp \lm} \right ) d^4 x = 0
\ee
corresponding to the GB theorem (Kobayashi and Nomizu 1969). In Eq. (3), 
Greek indices refer to four-dimensional coordinates. Introducing the GB 
term into the gra-vitational Lagrangian will not affect the four-dimensional 
field equations at all. However, within the context of the perturbative 
approach mentioned above, the most important contribution comes from the GB 
term (Mignemi and Stewart 1993). As in four dimensions it is a total 
divergence, to render this term dynamical, one has to consider a 
higher-dimesional background or to couple it to a scalar field.

The idea of a multi-dimensional spacetime has received much attention as a 
candidate for the unification of all fundamental interactions, including 
gravity, in the framework of supergravity and superstrings (Applequist 
et al 1987, Green et al 1987). In most higher-dimensional theories of gravity, 
the extra dimensions are assumed to form, at the present epoch, a compact 
manifold ({\em internal space}) of very small size compared to that of the 
three-dimensional visible space ({\em external space}) and therefore they are 
unobservable at the energies currently available (Green et al 1987). This 
so-called {\em compactification} of the extra dimensions may be achieved, in 
a natural way, by adding a square-curvature term $({\cal R}_{ijkl} 
{\cal R}^{ijkl})$ in the EH action of the gravitational field (M\"uller-Hoissen 
1988). In this way, the higher-dimensional theories are closely related to 
those of non-linear Lagrangians and their combination probably yields a natural 
generalization of GR.

In the present paper, we explore this generalization in view of the 
renormalizable energy-momentum tensor which acts as the source of gravity in 
the (semiclassical) interaction of the gravitational with a quantum matter 
field. In particular:

In Section II, we discuss briefly how GR is modified by the introduction 
of the renormalizable energy-momentum tensor first recognized by Calan et 
al (1970), on the rhs of the field equations. Introducing an analogous 
method, we explore the corresponding implications as regards a 
multi-dimensional higher-order gravity theory. We find that, in this case, 
the action functional, describing the semi-classical interaction of a quantum 
scalar field with the classical gravitational one, is being further modified 
and its variation with respect to the quantum field results in an inhomogeneous 
Klein-Gordon equation, the source term of which is purely geometric $( \sim 
{\cal R}^2)$. Accordingly, using a general method, we solve this equation, by 
means of the associated Green's function, in a compact, anisotropic 
six-dimensional background (Section III).

\section*{2. Quadratic Interaction}

Conventional gravity in $n-$dimensions implies that the dynamical behaviour 
of the gravitational field arises from an action principle involving the EH 
Lagrangian
\be
{\cal L}_{EH} = {1 \over 16 \pi G_n} \: {\cal R}
\ee
where $G_n = G V_{n-4}$ and $V_{n-4}$ denotes the volume of the internal space, 
formed by some extra spacelike dimensions. In this framework, we consider 
the semi-classical interaction between the gravitational and a massive quantum 
scalar field $\Phi (t, \vec{x})$ to the lowest order in $G_n$. The 
quantization of the field $\Phi (t, \vec{x})$ is performed by imposing 
canonical commutation relations on a hypersurface $t=$ constant (Isham 1981)
$$
\left [ \Ph (t, \vec{x}) \; , \; \Ph (t, {\vec{x}}^{\prime}) \right ] 
= 0 = \left [ \pi (t, \vec{x}) \; , \; \pi (t, {\vec{x}}^{\prime}) 
\right ]
$$
\be
\left [ \Ph (t, \vec{x}) \; , \; \pi (t, {\vec{x}}^{\prime}) 
\right ] = i \dl^{(n-1)} \left ( \vec{x} - {\vec{x}}^{\prime} \right ) 
\ee
where $\pi (t, \vec{x})$ is the momentum canonically conjugate to the 
field $\Ph(t, \vec{x})$. The equal-time commutation relations (5) guarantee 
the local character of the quantum field theory under consideration, thus 
attributing its time-evolution to the classical gravitational field equations 
(Birrell and Davies 1982). 

In any local field theory, the corresponding energy-momentum tensor is a very 
important object. Knowledge of its matrix elements is necessary to describe 
scattering in a relatively-weak external gravitational field. Therefore, in any 
quantum process in curved spacetime, it is desirable for the corresponding 
energy-momentum tensor to be {\em renormalizable}; i.e. its matrix elements to 
be cut-off independent (Birrell and Davies 1982). In this context, it has been 
proved (Callan et al 1970) that the functional form of the renormalizable 
energy-momentum tensor involved in the semi-classical interaction between the 
gravitational and a quantum field in $n-$dimensions, should be
\be
\Th_{ik} = T_{ik} - {1 \over 4} \: {n-2 \over n-1} \left [ {\Phi^2}_{; \: ik} - 
g_{ik} \Box \Phi^2 \right ]
\ee
where the semicolon stands for covariant differentiation $(\nabla_k)$, $\Box = 
g^{ik} \nabla_i \nabla_k$ is the d' Alembert operator and 
\be
T_{ik} = \Phi_{, i} \Phi_{, k} - g_{ik} {\cal L}_{mat}
\ee
is the {\em conventional} energy-momentum tensor of an (otherwise) free 
massive scalar field, with Lagrangian density of the form
\be
{\cal L}_{mat} = {1 \over 2} \left [ g^{ik} \Phi_{, i} \Phi_{, k} - m^2 
\Phi^2 \right ]
\ee
It's worthnoting that the tensor (6) defines the same $n-$momentum and Lorentz 
generators as the conventional energy-momentum tensor. 

It has been shown (Callan et al 1970) that the energy-momentum tensor (6) 
can be obtained by an action principle, involving 
\be
S = \int \left [ f(\Phi) {\cal R} + {\cal L}_{mat} \right ] \sqrt{- g} d^n x
\ee
where $f(\Phi)$ is an arbitrary, analytic function of $\Phi (t, \vec{x})$, 
the determination of which can be achieved by demanding that the rhs of the 
field equations resulting from Eq. (9) is given by Eq. (6). Accordingly, 
\be
{\dl S \over \dl g^{ik}} = 0 \Rightarrow {\cal R}_{ik} - {1 \over 2} g_{ik} 
{\cal R} = -8 \pi G_n \Th_{ik} = - {1 \over 2f} \left ( T_{ik} + 2 f_{; \: ik} 
- 2 g_{ik} \Box f \right )
\ee
To lowest order in $G_n$, one obtains (Callan et al 1970)
\be
f(\Phi) = {1 \over 16 \pi G_n} - {1 \over 8} {n-2 \over n-1} \Phi^2
\ee
Therefore, in any {\em linear Lagrangian gravity theory}, the interaction 
between a quantum scalar field and the classical gravitational one is 
determined through Hamilton's principle involving the action scalar 
\be
S = \int \sqrt {-g} \left [ ( {1 \over 16 \pi G_n} - {1 \over 8} 
{n-2 \over n-1} \Phi^2 ) \: {\cal R} + {\cal L}_{mat} \right ] d^n x
\ee

On the other hand, both superstring theories (Candelas et al 1985, Green et 
al 1987) and the one-loop approximation of quantum gravity (Kleidis and 
Papadopoulos 1998), suggest that the presence of quadratic terms in the 
gravitational action is {\em a priori} expected. Therefore, in connection 
to the semi-classical interaction previously stated, the question that 
arises now is, what the functional form of the corresponding {\em 
renormalizable} energy-momentum tensor might be, if the simplest quadratic 
curvature term, ${\cal R}^2$, is included in the description of the classical 
gravitational field. To answer this question, by analogy to Eq. (9), we may 
consider the action principle
\be
{\dl \over \dl g^{ik}} \int \sqrt {-g} \left [ f_1 (\Ph) {\cal R} + \al 
f_2 (\Ph) {\cal R}^2 + {\cal L}_{mat} \right ] d^n x = 0
\ee
where, both $f_1(\Ph)$ and $f_2(\Ph)$ are arbitrary, polynomial functions of 
$\Ph$. Eq. (13) yields
\be
{\cal R}_{ik} - {1 \over 2} g_{ik} {\cal R} = - {1 \over 2F} \left [ T_{ik} 
+ 2 F_{; \: ik} - 2 g_{ik} \Box F + \al g_{ik} f_2 (\Ph) {\cal R}^2 \right ]
\ee
where the function $F$ stands for the combination
\be
F = f_1 (\Ph) + 2 \al {\cal R} f_2 (\Ph)
\ee
For $\al = 0$ and to the lowest order in $G_n$ (but to every order in the 
coupling constants of the quantum field involved), we must have
\be
{\cal R}_{ik} - {1 \over 2} g_{ik} {\cal R} = -8 \pi G_n \Th_{ik}
\ee
where $\Th_{ik}$ [given by Eq. (6)] is the renormalizable energy-momentum 
tensor first recognized by Calan et al (1970). In this respect, we obtain 
$f_1 (\Ph) = f(\Ph)$, i.e. a function quadratic in $\Ph$ [see Eq. (11)]. 
Furthermore, on dimensional grounds regarding Eq. (14), we expect that
\be
F \sim \Ph^2
\ee
and therefore, $\al {\cal R} f_2(\Ph) \sim \Ph^2$, as well. However, we 
already know that ${\cal R} \sim [\Ph]$, as indicated by Whitt (1984), 
something that leads to $f_2 (\Ph) \sim \Ph$ {\footnote {In fact, 
$[{\cal R}] \sim [\Phi]^{4 \over n-2}$ and, therefore, $f_2 \sim \Phi^{2 
{n-4 \over n-2}}$. In order to render the coupling constant $\al$ 
dimensionless, one should consider $n=6$. Hence, $f_2 \sim \Phi$ only in 
six dimensions. The authors would like to thank the referee for pointing 
that out. }} and in particular, \be
F(\Phi) = {1 \over 16 \pi G_n} - {1 \over 8} {n-2 \over n-1} \left [ \Ph^2 
\right ] + 2 \al {\cal R} \Ph 
\ee
In Eq. (18), the coupling parameter $\al$ encapsulates any arbitrary constant 
that may be introduced in the definition of $f_2(\Ph)$. Accordingly, the 
action describing the semi-classical interaction of a quantum scalar field 
with the classical gravitational one up to the second order in curvature 
tensor, is being further modified and is written in the form
\be
S = \int \sqrt {-g} \left [ ( {1 \over 16 \pi G_n} - {1 \over 2} \ks_n \Ph^2 ) 
{\cal R} + \al {\cal R}^2 \Ph + {\cal L}_{mat} \right ] d^n x
\ee
where
\be
\ks_n = {1 \over 4} {n-2 \over n-1}
\ee
is the so-called {\em conformal coupling} parameter (Birrell and Davies 1982). 
In this case, the associated gravitational field equations (14) result in
\be
{\cal R}_{ik} - {1 \over 2} g_{ik} {\cal R} = -8 \pi G_n \: (\Th_{ik} + \al 
S_{ik})
\ee
where
\be
S_{ik} = g_{ik} \: {\cal R}^2 \Ph
\ee
The rhs of Eq. (21) represents the {\em "new"} renormalizable energy-momentum 
tensor. Notice that, as long as $\al \neq 0$, this tensor contains the extra 
{\em "source"} term $S_{ik}$. Inspite the presence of this term, the 
generalized energy-momentum tensor still remains renormalizable. This is 
due to the fact that, the set of the quantum operators $\lbrace \Phi, 
\Phi^2, \Box \Phi \rbrace$ is closed under renormalization, as it can be 
verified by straightforward power counting (see Callan et al 1970). 

Eq. (22) implies that the quadratic curvature term (i.e. pure global gravity) 
acts as a source of the quantum field $\Ph$. Indeed, variation of Eq. (19) 
with respect to $\Ph (t, \vec{x})$ leads to the following quantum field 
equation of propagation
\be
\Box \Ph + m^2 \Ph + \ks_n {\cal R} \Ph = \al {\cal R}^2
\ee
that is, an inhomogeneous Klein-Gordon equation in curved spacetime. In what 
follows we are interested in further exploring the interaction described by 
Eq. (23). It's worth pointing out that, in Eq. (19), the generalized coupling 
constant $\al$ remains dimensionless (and this is also the case for the 
corresponding action) only as long as 
\be
n = 6
\ee
thus indicating the appropriate spacetime dimensions for the semi-classical 
theory under consideration to hold, without introducing any additional 
arbitrary length scales. For this reason, in what follows, we restrict 
ourselves to a six-dimensional background.

\section*{3. A Green's Function in Six Dimensions}

In six dimensions, a self-consistent semi-classical coupling between 
${\cal R}^2$ and $\Ph (t, \vec {x})$ is introduced, leading to the 
"correct" renormalizable energy-momentum tensor which represents the source 
of gravity within the context of quadratic theories. Now, the corresponding 
action functional reads
\be
S = \int \sqrt {-g} \left [ ( {1 \over 16 \pi G_n} - {1 \over 10} \Ph^2 ) 
{\cal R} + \al {\cal R}^2 \Ph + {\cal L}_{mat} \right ] d^6 x
\ee
since $\ks_6 = {1 \over 5}$. Variation of this action with respect to the 
matter field, results in the following equation of motion
\be
{\dl S \over \dl \Ph} = 0 \Rightarrow \Box \Ph + m^2 \Ph + {1 \over 5} 
{\cal R} \Ph = \al {\cal R}^2
\ee
Eq. (26) describes the propagation of an (otherwise free) real quantum 
scalar field in a six-dimensional curved background in the presence of 
a geometric source $({\cal R}^2)$. To solve this equation with respect 
to $\Ph$ is a quite delicate problem. One has to seek for solutions in 
the form
\be
\Ph (x) = - \al \: \int d^{\: 6} x^{\prime} \sqrt {-g (x^{\prime})} 
G (x, x^{\prime}) {\cal R}^2 (x^{\prime})
\ee
that is, to solve the following differential equation for the Green's function 
$G (x , x^{\prime} )$ in a six-dimensional spacetime
\be
\left [ \Box + m^2 + {1 \over 5} {\cal R} \right ] G(x, x^{\prime}) = {1 
\over {\sqrt{-g (x)}}} \dl^{(6)} (x - x^{\prime})
\ee
In order to do so, we have to determine an appropriate model Universe. In 
view of most higher-dimensional theories of gravity (Applequist et al 1987), 
we consider a six-dimensional background in which both the external and the 
internal space represent compact manifolds of constant curvature (i.e. $k_{ext} 
= + 1$ and $k_{int} = + 1$). Accordingly, the metric of the six-dimensional 
spacetime reads
\be
ds^2 = -dt^2 + e^{2 \om(t)} \left [ d \ch^2 + sin^2 \ch (d \th^2 + sin^2 \th 
d \ph^2) \right ] + e^{2 w(t)} \left [ d \ps^2 + sin^2 \ps d \ups^2 
\right ]
\ee
where the (logarithmic) scale functions of the external and the internal space 
$\om (t)$ and $w(t)$ respectively, depend on the time-coordinate only. 
Due to the compactness of the spatial part, the Green's function under 
consideration is subject to spatially periodic boundary conditions. Furthermore, 
in a spatially homogeneous spacetime of the form (29), we can always perform 
the translation $x - x^{\prime} \rightarrow x$ (i.e. $x^{\prime} = 0$). In 
this case, the action of the wave operator $\Box = g^{ik} \nabla_i \nabla_k$ 
on $G(x)$ reads
\bea
g^{ik} \nabla_i \nabla_k G (x) = &-&
[G_{,tt}+(3\dot{\om }+2\dot{w})G_{,t}]+ \nn \\
&+&e^{-2\om }[G_{,\chi \chi }+
\frac{2cos\chi }{sin\chi }G_{,\chi }]+\nn \\
&+&\frac{e^{-2\om }}{sin^{2}\chi }
[G_{,\th \th }+\frac{cos\th }{sin\th }G_{,\th }]+
\frac{e^{-2\om }}{sin^{2}\chi sin^{2}\th }G_{,\phi \phi }+ \nn \\
&+&e^{-2w}[G_{,\psi \psi }+
\frac{cos\psi }{sin\psi }G_{,\psi }]+
\frac{e^{-2w}}{sin^{2}\psi }G_{,\ups \ups }
\eea
We propose that any linear combination of solutions in the form
\bea
G (x) = G_{i}(t)
\frac{P_{L+1/2}^{l+1/2}(cos\chi )}{\sqrt{sin\chi }}
Y_{l}^{m}(\th ,\phi )
Y_{\bar{l}}^{\bar{m}}(\psi ,\ups )
\; \; \; \; ( i = \lbrace L, l, \bar{l}, m, \bar{m} \rbrace )
\eea
will render the wave operator time-dependent only. We consider the case that, 
$L\geq l\geq 0$ are non-negative integers so the functions $P_{L+1/2}^{l+1/2}
(cos \chi)$, are polynomials (see Appendix I). As $l$ is kept constant while 
$L$ increases, one obtains a full set of polynomials of degree $(L-l)$. By 
direct substitution into Eq. (A1) it can be shown that for the even-degree 
polynomials one obtains $(z = cos \chi)$
\be
P_{L+1/2}^{l+1/2}(z)=\sum_{\lm =0}^{k^{*}}c_{2\lm } z^{2\lm }
\; \; \; \; \; (2k^{*}\equiv L-l)
\ee
where
\be
c_{2\lm }=\frac{2^{2\lm }(-1)^{\lm }}{(2\lm )!}
\frac{(k^{*})!}{(k^{*}-\lm )!}
\frac{(\lm +k^{*}+l)!}{(k^{*}+l)!}c_{0}\; \; \; \; \;
(0\leq \lm \leq k^{*})
\ee
In this case, both $l\geq 0$ and $c_{0}$ are constants. On the other hand, 
for the odd-degree polynomials we have
\be
P_{L+1/2}^{l+1/2}(z)=\sum_{\lm =0}^{k^{*}}c_{2\lm +1} z^{2\lm +1}
\; \; \; \; \; (2k^{*}+1\equiv L-l)
\ee
where
\be
c_{2\lm +1}=\frac{2^{2\lm }(-1)^{\lm }}{(2\lm +1)!}
\frac{(k^{*})!}{(k^{*}-\lm )!}
\frac{(\lm +k^{*}+l+1)!}{(k^{*}+l+1)!}c_{1}\; \; \; \; \;
(0\leq \lm \leq k^{*})
\ee
In fact, both Eqs. (32) and (34) represent the Gegenbauer (Ultraspherical) 
polynomials (see Abramowitz and Stegun 1970, p. 774-775, 22.2.3, 22.3.4). 
Indeed, substitution of Eq. (31) [through Eq. (30)] into Eq. (28) will render 
$P_{L+1/2}^{l+1/2}(cos\chi )$ satisfying Eq. (A1) with the appropriate degree 
and order, provided that the function $G_i (t) = G(t)$ satisfies
\bea
-G_{,tt} &-& (3\dot{\om } + 2\dot{w}) G_{,t}
-L(L+2) e^{-2\om }G - \bar{l} (\bar{l}+1) e^{-2w} G + \nn \\
&+& m^{2} G + {1 \over 5} {\cal R}(t) G = 0
\eea
Then, the Green's function which yields an appropriate subclass of 
solutions to Eq. (26), through Eq. (27), is given by 
\be
G = \sum_{i} g_i G_i (t)
\frac{C_{L-l}^{(l+1)}(cos\chi )}{\sqrt{sin\chi }}
Y_{l}^{m}(\th,\phi )Y_{\bar{l}}^{\bar{m}}(\psi ,\ups )\; \; \; \;
( i = \lbrace L, l, \bar{l}, m, \bar{m} \rbrace )
\ee
where $g_i$ are arbitrary expansion coefficients, to be determined by the 
boundary conditions. As regards the time-evolution of the modes, Eq. (36) 
can be written in the form
\be
{d \over dt} \left [ e^{3 \om + 2 w} {d \over dt} G \right ] - \left [ 
m^2 + \Om (t) \right ] e^{3 \om + 2 w} \: G = 0
\ee
where, we have set
\be
\Om (t) = {1 \over 5} {\cal R} (t) - L(L+2) e^{-2\om (t) } - \bar{l} 
(\bar{l}+1) e^{-2w(t)}
\ee
Defining the so-called {\em conformal time} as
\be
d \et = e^{-(3 \om + 2 w)} \: dt
\ee
we obtain
\be
{d^2 \over d \et^2} G - \left [ m^2 + \Om (\et) \right ] G = 0
\ee
Now, provided that $m \neq 0$ and defining the dimensionless time-parameter 
$\ks$ as
\be
\ks = m \et = m \: \int^t {dt \over e^{3 \om + 2 w}}
\ee
we, finally, obtain
\be
{d^2 \over d \ks^2} G - \left [ 1 + \varep (\ks) \right ] G = 0
\ee
where we have set
\be
\varep (\ks) = {1\over m^2} \: \Om(\ks) = {1 \over m^2} \: \left [ 
{1 \over 5} {\cal R} (\ks) - L(L+2) e^{-2\om (\ks) } - \bar{l} (\bar{l}+1) 
e^{-2w (\ks)} \right ]
\ee
Solutions to Eq. (43) exist, and can be put in the formal form
\be
G(\ks) = e^{\ks} \left [ 1 + h(\ks) \right ]
\ee
To derive the functional form of $h(\ks)$, we combine Eqs. (43) and 
(45), thus obtaining
\be
h^{\prime \prime} (\ks) + 2 h^{\prime} (\ks) - \varep (\ks) h(\ks) = 
\varep (\ks)
\ee
where a prime denotes differentiation with respect to $\ks$. To solve this 
inhomogeneous differential equation for $h (\ks)$, we apply the method of 
{\em variation of parameters} (Davis 1962). The result can be written as 
an integral equation of Volterra's type
\be
h(\ks) = {1 \over 2} \int_0^{\ks} \left [ 1 - e^{2 (\zt - \ks)} \right ] 
\varep (\zt) \left [ 1 + h (\zt) \right ] d \zt
\ee
Conversely, it can be verified by differentiation that, any 
twice-differentiable solution of Eq. (47) satisfies Eq. (46). 
Eq. (47) can be solved by a method of {\em succes-sive approximations} 
(Olver 1974). Accordingly, we define the sequence $h_s(\ks)$ 
$(s = 0, 1, 2, 3, ...)$, with 
\be
h_0 (\ks) = 0
\ee
and
\be
h_s(\ks) = {1 \over 2} \int_0^{\ks} \left [ 1 - e^{2 (\zt - \ks)} \right ] 
\varep (\zt) \left [ 1 + h_{s-1} (\zt) \right ] d \zt
\ee
$(s \geq 1)$ where, in particular,
\be
h_1(\ks) = {1 \over 2} \int_0^{\ks} \left [ 1 - e^{2 (\zt - \ks)} \right ] 
\varep (\zt) d \zt
\ee
Since $\ks - \zt \geq 0$, we have $0 \leq 1 - e^{2 (\zt - \ks)} < 1$ and 
therefore
\be
\vert h_1 (\ks) \vert \leq {\Ps (\ks) \over 2}
\ee
where
\be
\Ps (\ks) = \int_0^{\ks} \vert \varep (\ks) \vert d \zt
\ee
and the equality in Eq. (51) holds only for $\ks = 0$. Then, by induction, 
we find that the relation
\be
\vert h_s (\ks) - h_{s-1} (\ks) \vert \leq {\Ps^s (\ks) \over 2^s \: s!}
\ee
also holds, for every $s$. Now, as long as $\Ps (\ks)$ is bounded, the series
\be
h(\ks) = \sum_{s=0}^{\infty} [ h_{s+1} (\ks) - h_s (\ks) ]
\ee
converges uniformly in any compact $\ks$ interval. Then, by summation of Eq. 
(54) and the use of Eqs. (49) and (50) we verify that $h(\ks)$ satisfies the 
integral equation (47). Therefore, the general solution to Eq. (43) is written 
in the form
\be
G(m \et) = e^{m \et} \left ( 1 + \sum_{s=0}^{\infty} [h_{s+1} (m \et) - 
h_s (m \et)] \right )
\ee
with $h_s(m \et)$ being given by Eq. (49). Finally, the combination of Eqs. 
(37) and (55) results in the Green's function associated with a subclass of 
solutions of Eq. (26), through Eq. (27).

\vspace{.5cm}

Summarizing, a self-consistent coupling between the square curvature term 
${\cal R}^2$ and the quantum field $\Ph (t, \vec{x})$ should be introduced 
in order to yield the "correct" renormalizable energy-momentum tensor in 
non-linear gravity theories. The subsequent {\em "quadratic interaction"} 
discards any higher-order derivative terms from the gravitational field 
equations, but it introduces a {\em geometric source} term in the wave 
equation for the quantum field. In this case, unlike the {\em conventional} 
conformal coupling $(\sim {\cal R} \Ph^2)$, the quantum field interacts with 
gravity not only through its mass (or energy) content $(\sim \Ph^2)$, but, 
also, in a more generic way $({\cal R}^2 \Ph)$. Within this context, we 
propose a general method of obtaining mode-solutions for the quantum field 
[Eq. (27)], by means of the associated Green's function [Eqs. (37) and (55)], 
in an anisotropic six-dimensional background.

\section*{Appendix I}

The Legendre functions, satisfying the relation 
$$ 
\frac{d}{dz}
\left[(1-z^{2})\frac{dP_{l}^{m}(z)}{dz}\right]+
\left[l(l+1)-\frac{m^{2}}{1-z^{2}}\right]P_{l}^{m}(z)=0\; \; \; \;
(z = cos \ch) \eqno{(A1)}
$$ 
can be expanded as (e.g. see Abramowitz and Stegun 1970)
$$ 
P_{n}^{m}(z)=\sum_{l=l_{min}}^{n-m} c_{l} z^{l}\; \; \; \; (l_{min}=0,1) 
\eqno{(A2)}
$$ 
where 
$$
\frac{c_{l+2}}{c_{l}}=\frac{(l+m)(l+m+1)-n(n+1)}{(l+1)(l+2)} \eqno{(A3)}
$$ 
and therefore, they are polynomials as long as $n-m \geq 0$. Upon 
the correspondence $n\rightarrow 2k^{*}$, $n-2m\rightarrow 
2\lm $, $m\rightarrow k^{*}-\lm $ and setting 
$$ 
c_{0}=\frac{(-1)^{k^{*}}}{l!}\frac{(k^{*}+l)!}{(k^{*})!} \eqno{(A4)}
$$ 
we find that, in the even-degree case,
$$ 
P_{L+1/2}^{l+1/2}(z) = C_{2k^{*}}^{(l+1)}(z) \; \; \; \; (2k^{*}=L-l) 
\eqno{(A5)}
$$ 
On the other hand, for the odd-degree case, performing the correspondence 
$n\rightarrow 2k^{*}+1$, $n-2m\rightarrow 2\lm +1$, 
$m\rightarrow k^{*}-\lm $ and setting the normalization constant 
$c_1$ as
$$ 
c_{1}=\frac{(-1)^{k^{*}}}{l!}
\frac{(k^{*}+l+1)!}{(k^{*})!} \eqno{(A6)}
$$ 
one, accordingly, obtains 
$$ 
P_{L+1/2}^{l+1/2}(z) = C_{2k^{*}+1}^{(l+1)}(z) \; \; \; \; 
(2k^{*}+1=L-l) \eqno{(A7)}
$$ 

\section*{References}

\begin{itemize}

\item[]Abramowitz M and Stegun A I 1970 {\em Handbook of Mathematical 
Functions} (New York: Dover)

\item[]Applequist T, Chodos A and Freund P 1987 {\em Modern Kaluza-Klein 
Theories} (Menlo Park, CA: Addison-Wesley)

\item[]Birrell N D and Davies P C W 1982 {\em Quantum Fields in Curved Space} 
(Cambridge: Cambridge University Press)

\item[]Calan C G, Coleman S and Jackiw R 1970 {\it Ann. Phys.} {\bf 59} 42

\item[]Candelas P, Horowitz G T, Strominger A and Witten E 1985 {\it Nucl. 
Phys. B} {\bf 258} 46

\item[]Davis F 1962 {\em Introduction to Non-linear Differential and Integral 
Equations} (New York: Dover)

\item[]Green M B, Schwartz J H and Witten E 1987 {\em Superstring Theory} 
(Cambridge: Cambridge University Press)

\item[]Isham C J 1981 Quantum Gravity - an overview {\em Quantum Gravity: a 
Second Oxford Symposium} ed C J Isham, R Penrose and D W Sciama (Oxford: 
Clarendon)

\item[]Kleidis K and Papadopoulos D B 1998 {\it Class. Quantum Grav.} {\bf 15} 
2217

\item[]Kobayashi S and Nomizu K 1969 {\em Foundations of Differential Geometry 
II} (New York: Wiley InterScience)

\item[]Lovelock D 1971 {\it J. Math. Phys.} {\bf 12} 498

\item[]Mignemi S and Stewart N R 1993 {\it Phys. Rev. D} {\bf 47} 5259

\item[]M\"uller-Hoissen F 1988 {\it Class Quantum Grav.} {\bf 5} L35

\item[]Nanopoulos D B 1997 {\it preprint} hep-th/9711080

\item[]Olver F W 1974 {\em Asymptotics and Special Functions} (New York: 
Academic Press)

\item[]Polchinski J 1998 {\em String Theory} (Cambridge: Cambridge University 
Press)

\item[]Padmanabhan T 1989 {\it Int. J. Mod. Phys. A} {\bf 18} 4735

\item[]Schwartz J H 1999 {\it Phys. Rep.} {\bf 315} 107

\item[] Whitt B 1984 {\it Phys. Lett. B} {\bf 145} 176

\end{itemize}

\end{document}